\documentstyle[aps,12pt]{revtex}

\begin{document}
\title{The 3$d$-electron states in {\it \ }LaMnO$_{3}^{\ast }$}
\author{R.J. Radwanski}
\address{Center for Solid State Physics, S$^{nt}$ Filip 5,31-150Krakow,Poland%
\\
Institute of Physics, Pedagogical University, 30-084 Krakow, Poland}
\author{Z. Ropka}
\address{Center for Solid State Physics, S$^{nt}$ Filip 5,31-150Krakow,Poland%
\\
email: sfradwan@cyf-kr.edu.pl, http://www.css-physics.edu.pl}
\maketitle

\begin{abstract}
Three fundamentally different electronic structures for 3$d$ electron states
in LaMnO$_{3}$, discussed in the current literature, have been presented. We
are in favour of the localized electron atomic-like crystal-field approach
that yields the discrete energy spectrum, in the scale of 1 meV, associated
with the atomic-like states of the Mn$^{3+}$ ions in contrary to the
continuous energy spectrum yielded by band theories. In our atomic-like
approach the {\it d} electrons form the highly-correlated electron system 3$%
d^{n}$ described by two Hund's rules quantum numbers $S$=2 and $L$=2. We
take into account the spin-orbit coupling, Jahn-Teller distortion and
formation of the magnetic state. The resulting electronic structure is
completely different from that presented in the current literature. The
superiority of our model relies in the fact that it explains consistently
properties of LaMnO$_{3}$, the insulating and magnetic ground state and
thermodynamics, using well-established physical concepts.

PACS\ No: 71.70.Ej : 75.10.Dg : 73.30.-m;

\end{abstract}

\date{(Receipt: 19 December 2001)}

LaMnO$_{3}$ belong to the class of compounds known as Mott insulators. It
attracts the scientific interest by more than 50 years. The interest to LaMnO%
$_{3}$-based compounds has increased considerably after revealing its
colossal magnetoresistance properties \cite{1,2}. Despite of enormous
activities, both theoretical and experimental, understanding of the colossal
magnetoresistance and, related to it, of the magnetism and of the electronic
structure is still under very strong debate. In fact, the understanding of
the magnetism and of the electronic structure is very closely related to a
general problem ''how to treat $d$ electrons in 3$d$-atom containing
compounds''. The fundamental controversy ''how to treat $d$ electrons''
starts already at the beginning - should they be treated as localized or
itinerant (in the scientific physics sense, i.e. the localized or the
itinerant picture should be taken as the start). Directly related to this
problem is the structure of the available states: do they form the
continuous energy spectrum like it is in the band picture or do they form
the discrete energy spectrum typical for the localized states.

The aim of this paper is to show up the fundamentally different descriptions
of the electronic structure and the magnetism of LaMnO$_{3}$ appearing in
the current literature in prestigious physical journals by presenting 3
different electronic structures for the 3$d$ electrons. The one is related
to the band description and other two to the localized description. There is
a number of band-structure calculations, based on different versions of the
local density (LDA, LSDA) \cite{3,4,5,6,7} and Hartree-Fock approximations
(HFA) \cite{8,9}. The band results are schematically shown in Fig. 1. The
LDA/LSDA results are slightly different from the HFA\ results as far as the
position of Mn 3$d$ states with respect to the Fermi level and the nature of
states at the Fermi level are considered but both yield continuous energy
spectrum for the 3$d$ states spread over $\sim $10 eV. A gap, needed for the
insulating state, within the band picture is obtained by the Jahn-Teller
(JT) distortion \cite{5}.

In the current literature, apart from the band picture, often appears a
localized electronic structure \cite{10,11,12,13}. Some representative
results are presented in Figs 2-4. According to this picture 4 electrons
(spins) of the Mn$^{3+}$ ion are put subsequently one by one on the
single-electron states formed for one $d$ electron in the octahedral crystal
field (CEF): three spins are put to $t_{2g}$ orbitals and one occupies one
of the higher $e_{g}$ orbitals. Owing to only one electron in the \ $e_{g}$
doublet one can say that $e_{g}$ is half filled and the $e_{g}$ electron has
got the orbital degree of freedom. According to Refs \cite{10,11,12,13} such
the electron/spin arrangement satisfies the first Hund's rule. In this
approach the $e_{g}$ electron plays the main role in the further description
of properties of manganites.

The general shape of the bands presented in Fig. 1 can be somehow understood
knowing the localized states of Figs 2-4. The continuous energy spectrum
looks like the smooth convolution on the available localized single 3$d$
electron orbitals $t_{2g}$ (fully occupied) and higher $e_{g}$ orbitals
(half-filled). The similarity of the band density of states and the energy
level scheme of Figs 2-4 is related, according to us, to the single-electron
treatment of the 3$d$ electrons in both approaches.

The third type of the electronic structure, derived by us within the
localized picture, Refs \cite{14,15}, but with taking into account two
Hund's rules and the spin-orbit coupling, is presented in Fig. 5. In this
atomic-like CEF approach the 4 {\it d} electrons in the open 3{\it d} shell
of the Mn$^{3+}$ ion form the highly-correlated \ intra-atomic 3$d^{4}$
electron system. These strong correlations among the 3$d$ electrons are
accounted for, in a zero-order approximation, by two Hund's rules. These two
Hund's rules yield for the $d^{4}$ system the ground term $^{5}D$ with $S$=2
and $L$=2. Its 25-fold degeneracy is removed by the crystal field and
spin-orbit interactions. Under the action of the dominant cubic crystal
field, the $^{5}D$ term splits into the orbital triplet $^{5}T_{2g}$ and the
orbital doublet $^{5}E_{g}$ with the energy separation $\Delta $ of about
1-3 eV. In the octahedral oxygen-anion surrounding of the Mn$^{3+}$ ion,
realized in the perovskite-like structure of LaMnO$_{3}$, the orbital
doublet $^{5}E_{g}$ is lower. The $E_{g}$ ground subterm is confirmed by our 
{\it ab initio} calculations of the hexadecapolar potential (the $A_{4}$ CEF
coefficient) at the Mn site in the oxygen-anion octahedron. The $A_{4}$ CEF
coefficient in the center of the oxygen-anion octahedron is positive. The $%
A_{4}$ CEF coefficient together with the Stevens factor $\beta $ for the 3$%
d^{4}$ system, of -2/63 (\cite{16}, table 19 on page 873), yields the
negative value of the octahedral CEF parameter $B_{4}$ and consequently the $%
^{5}E_{g}$ ground subterm. The negative sign for $B_{4}$ has been derived
already a pretty long time ago by Abragam and Bleaney (\cite{16}, see page
374). However, this result has been completely forgotten in the current
physics as in the current discussion of 3$d$-atom compounds figures like
those presented in Figs 2-4 with the $t_{2g}$ ground orbitals and excited $%
e_{g}$ orbitals are always considered.

The low-energy electronic structure has been calculated by us from the
single-ion-like Hamiltonian, considered within the 25-fold spin-orbital
space (the $\left| LSL_{\text{z}}S_{\text{z}}\right\rangle $ space),
containing simultaneously the crystal-field (of the optional symmetry) and
the spin-orbit coupling. The structure shown in Fig. 5 is obtained for the
purely octahedral symmetry but in the presence of the spin-orbit coupling.
The splitting of the $^{5}E_{g}$ subterm is the spin-orbit effect.

In the reality Nature with the decreasing temperature involves further
effects - a lattice distortion, magnetic interactions, that often lead to
the magnetic state, and/or a site differentiation with respect to the local
surroundings and the local symmetry. All these effects are observed in LaMnO$%
_{3}$, indeed and they are computable in our approach. In general, all these
effects cause the subsequent lifting of the degeneracy of states visible in
Fig. 5 as well as the lowering of the energy of the ground state and a quite
substantial spreading of 10 lowest states.

The tetragonal off-octahedral distortion, relevant to the situation realized
in LaMnO$_{3}$, decreases the energy of the ground state. Thus, the
distortion can go spontaneously and satisfies the Jahn-Teller theorem, like
we have calculated for LaCoO$_{3}$ \cite{17}.

In the magnetic state the magnetic inter-site interactions prevail the
temperature disordering. We have performed self-consistent calculations
considering the magnetic interactions within the molecular-field
approximation for two cases, for the magnetic structure with the magnetic
moments along the tetragonal $c$ axis and for moments along the $a$ axis.
The same value of the molecular-field coefficient $n$\ of 17.7 K/$\mu
_{B}^{2}$ ( =26.35 T/$\mu _{B}$) produces the magnetic order below 95 K for
moments along the tetragonal $c$ axis and below 140 K for the magnetic order
with moments along $a$ axis. We interpret this theoretical result as the
preference of the system to order magnetically with moments along the $a$
axis. If so, it is in nice agreement with the experimental observation. In
the magnetic state a molecular field is set up self-consistently. Its value
is calculated to be 98 T at zero temperature. The electronic structure in
the magnetic state, shown in Fig. 6, and its temperature dependence can be
experimentally verified as it affects many physical properties like
temperature dependence of the heat capacity, of the paramagnetic
susceptibility $\chi $, of the magnetic-moment value. The energy separations
should be detected by energy spectroscopy experiments like Raman
spectroscopy, for instance. In fact, we take the results of the Raman
spectroscopy \cite{18} revealing a number of well-defined energy excitations
as a confirmation of our atomistic model \cite{19}. The obtained value of
the ordered magnetic moment at zero temperature amounts to 3.72 $\mu _{B}$.
It is built up from the spin moment of +3.96 $\mu _{B}$ and from the orbital
moment of -0.24 $\mu _{B}$. The opposite sign of the spin and orbital
moments mimics somehow the 3$^{rd}$ Hund rule. We point out that the
opposite sign of the spin and orbital moments comes out directly in our
calculations and is related to the positive sign of the intra-atomic
spin-orbit coupling. The effective moment calculated from the $\chi ^{-1}$ 
{\it vs} T plot in the 300-400 K region amounts to 4.65 $\mu _{B}$. It is
only slightly smaller value than the spin-only value of 4.90 $\mu _{B}$ what
is a surprise owing to the fact that we take into calculations the full
quantum orbital value $L$.

Our description of the electronic structure of LaMnO$_{3}$ proceeds within a
relatively weak crystal-field regime in contrary to the current-literature
view that prefers the strong crystal-field regime in the description of 3$d$%
-atom compounds. According to our picture, the crystal field is \ weak in
the sense that it does not break the intra-atomic arrangement of electrons
within the 3$d$ shell. Our approach is in sharp contrast to the strong CEF\
regime that is the basis for the single-electron description discussed
previously. It means, that we think that the intra-atomic structure of a
paramagnetic atom largely perseveres even when this atom becomes the full
part of a solid - on this basis we have developed a Quantum Atomistic
Solid-State Theory (QUASST) for compounds containing open 3$d$-/4$f$-/5$f$%
-shell atoms \cite{20}. This weak crystal-field approach, known within the
rare-earth CEF\ community as the CEF approach, has been often successfully
applied to 3$d$-ion doped systems, when 3$d$ ions were introduced as
impurities, in interpretation of, for instance,
electron-paramagnetic-resonance experiments \cite{16,21,22}. In Refs \cite%
{17,23,24,25} we have applied this approach to a 3$d$-ion system where\
Co/Fe/Ni ions are the full part of a solid forming LaCoO$_{3}$, FeBr$_{2}$
or NiO. Of course, in our picture the crystal field is much stronger than
the spin-orbit coupling as is generally accepted in the 3$d$ magnetism -
thus we work in the so-called intermediate crystal-field limit.

We would like to point out that our approach should not be considered as the
treatment of an isolated ion - we consider the cation in the octahedral
crystal field. This octahedral crystal field is predominantly associated
with the oxygen octahedron MnO$_{6}$. The perovskite structure is built up
from the corner sharing octahedra MnO$_{6}$ - thus such the atomic structure
occurs at each cation due to the translational symmetry. The strength of the
crystal field interactions is determined by the whole charge surroundings,
not only by the nearest oxygen octahedron. It makes that the CEF approach
looks like a single-ion approach but in fact it describes the coherent
states of the whole crystal. The parameters used are fully physical: the
octahedral CEF parameter B$_{4}$ =-13 meV, the tetragonal B$_{2}^{0}$ = +10
meV, the orthorhombic B$_{2}^{2}$ = - 2 meV and the spin-orbit coupling $%
\lambda $= +33 meV. They yield overall effect of 1.6, 0.10, 0.002 and 0.33
eV, respectively.

In conclusion, we have presented three fundamentally different electronic
structures for the 3$d$-electron states in LaMnO$_{3}$ discussed in the
current literature. We are in favour of the localized electron atomic-like
crystal-field approach that yields the discrete energy spectrum associated
with the atomic-like states of the Mn$^{3+}$ ions. The resulting electronic
structure is completely different from those presented in the current
literature. Our CEF-like model takes into account very strong correlations
within 3{\it d} electrons that form highly-correlated intra-atomic electron
system 3$d^{4}$. These strong correlations are accounted for by two Hund's
rules yielding $S$=2 and $L$=2. Our studies have revealed strong
correlations between the local magnetic moment (its value and the direction)
and the local distortions (more generally - the local symmetry of the
crystal field). Our approach provides in the very natural way the insulating
ground state for LaMnO$_{3}$ independently on the lattice distortions. The
superiority of our model relies in a fact that it consistently describes
both zero-temperature properties of LaMnO$_{3}$ and thermodynamics and it
makes use of the well-defined physical concepts. Good description of many
electronic and magnetic properties within the many-electron CEF model
indicates that the band-structure calculations have to be oriented into the
very strong intra-atomic {\it d}-{\it d} correlation limit in order to get
the ground state in agreement with {\bf two} Hund's rules and to take into
account the orbital magnetism.

* This paper has been submitted to Phys.Rev.Lett. on 19.12.2001, LM8412, but
has been rejected by the Editor as ''not suitable'' despite of our extensive
arguments that the open scientific discussion of the electronic structure
and magnetism of LaMnO$_{3}$ is very important in particular in the
situation when the electronic structure and properties of LaMnO$_{3}$ are
widely discussed in many prestigious physical journals and we came out with
novel electronic structures. Moreover, we argued that such arbitrary
decisions of the Editor of Phys.Rev.Lett. violate fundamental scientific
rules.

{\bf Figure caption:}

Fig. 1. Schematic description of the $d$ states in LaMnO$_{3}$ within the
band approach obtained within the LSDA approach - there is the continuous
energy spectrum. There is also some contribution from 2$p$ states of oxygen.
A slightly different picture is obtained within the Hartree-Fock
approximation, but still $d$ states form the continuous energy spectrum
spread over 10 eV.

Fig. 2. Electronic structure of a) Mn$^{3+}$ in octahedral coordination,
before and after Jahn-Teller distortion; b) Mn$^{4+}$ in octahedral
coordination; c) ferromagnetic manganites with x=0.3. After Ref. \cite{10}.

Fig. 3. Splitting of the five-fold degenerate atomic 3$d$ levels into lower $%
t_{2g}$ and higher $e_{g}$ states. The particular Jahn-Teller distortion
further lifts each degeneracy as shown. After \cite{11,12}.

Fig. 4. Energy level scheme for LaMnO$_{3}$ shown as Fig. 1 of Ref. \cite{2}%
. Following Ref. \cite{2} the values of E$_{x}$, E$_{cf}$ and E$_{JT}$ from
density functional calculations are around 3 eV, 2 eV and 1.5 eV
respectively. Typical bandwidth is 1.5-2.0 eV. The Fermi level lies between
JT split orbitals.

Fig. 5. The fine electronic structure of the Mn$^{3+}$ ion in the MnO$_{6}$
octahedron, realized in LaMnO$_{3}$, produced by the octahedral crystal
field and the intra-atomic spin-orbit coupling. The degeneracy and the
magnetic moment of the states are shown. The fine splitting is not to the
left-hand scale.

Fig. 6. The lowest part of the fine electronic structure of the Mn$^{3+}$
ion in LaMnO$_{3}$ in the magnetic state, below 140 K, with the magnetic
moments along the $a$ axis. Only 5 states originating from the $^{5}E_{g}$
subterm are shown, other 5 are about 100 meV above and show quite similar
behaviour. This splitting is mainly determined by the tetragonal distortion.
The small difference in the electronic structure in the paramagnetic region
in comparison to Fig. 5 is the effect of distortions.

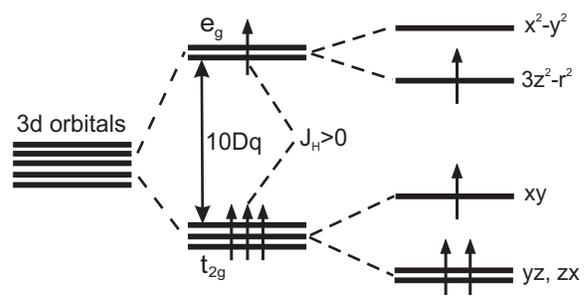
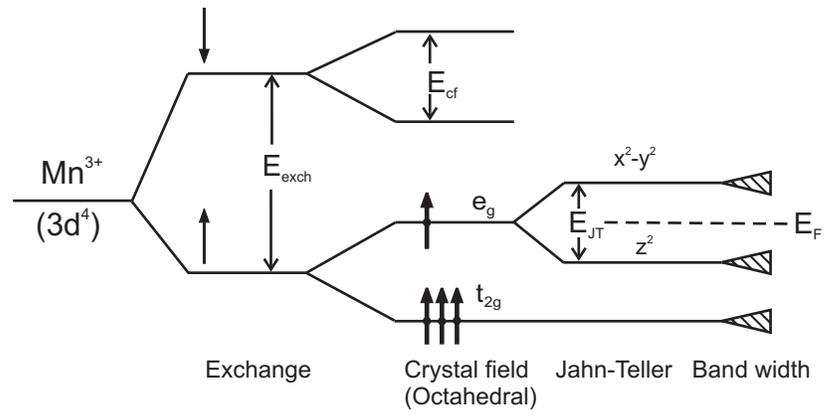
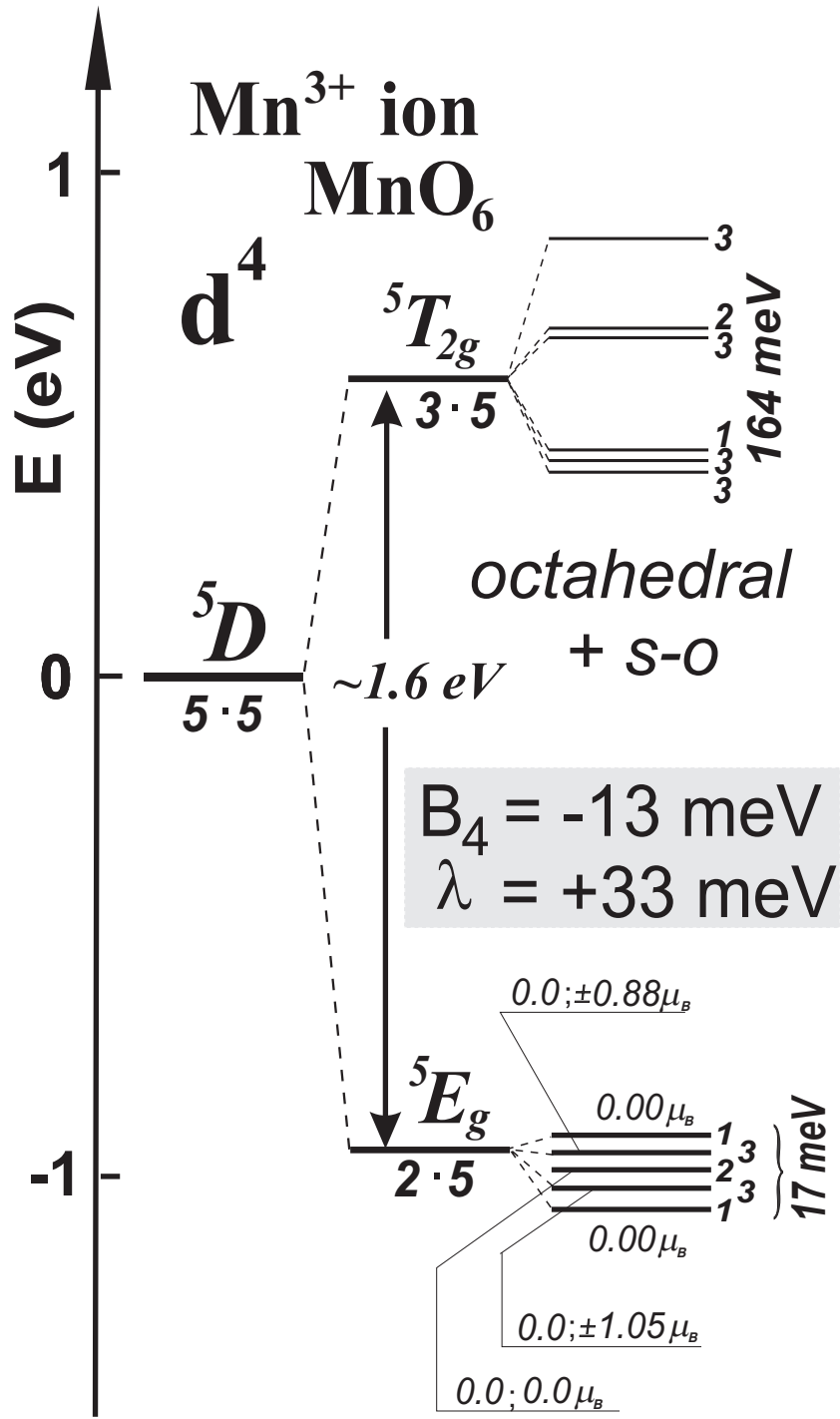
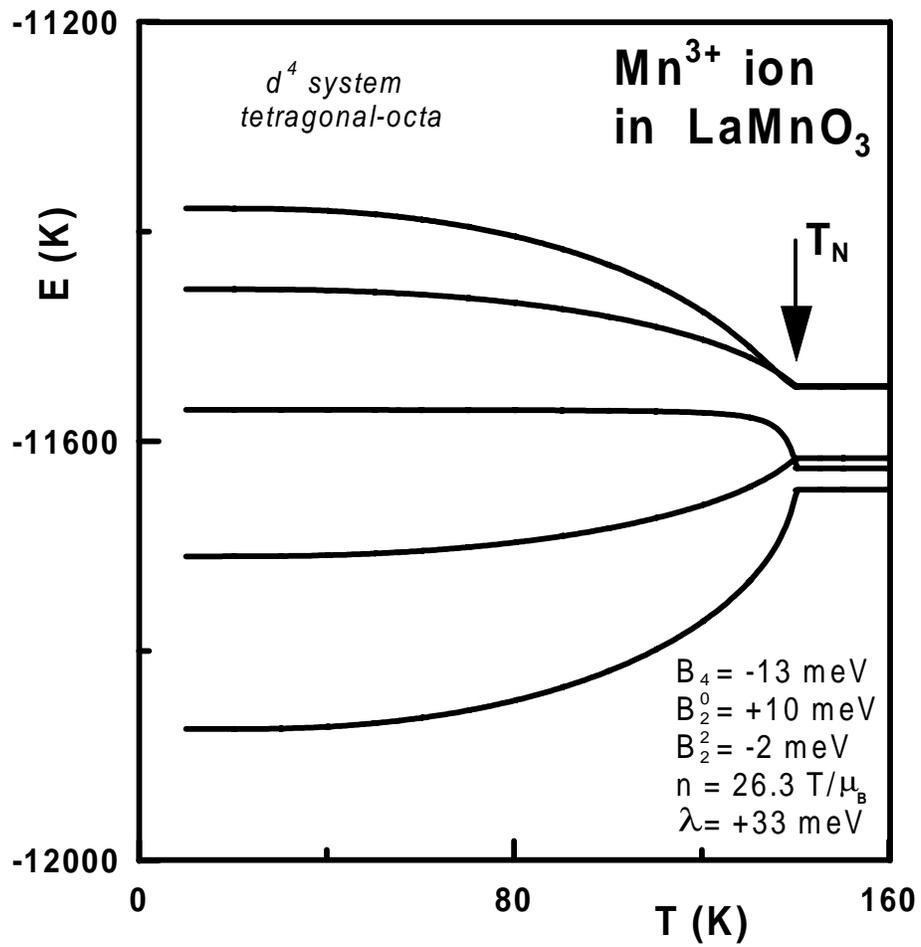
\end{document}